\documentclass[conference]{IEEEtran}
\IEEEoverridecommandlockouts
\usepackage{cite}
\usepackage{amsmath,amssymb,amsfonts}
\usepackage{algorithmic}
\usepackage{graphicx}
\usepackage{textcomp}
\usepackage{xcolor}
\usepackage{comment}
\def\BibTeX{{\rm B\kern-.05em{\sc i\kern-.025em b}\kern-.08em
    T\kern-.1667em\lower.7ex\hbox{E}\kern-.125emX}}
\usepackage{fancyhdr}

\begin{document}

\title{Two New CNTFET Quaternary Full Adders for Carry-Propagate Adders\\
%{\footnotesize \textsuperscript{*}Note: Sub-titles are not captured in Xplore and
%should not be used}
%\thanks{Identify applicable funding agency here. If none, delete this.}
}

\author{\IEEEauthorblockN{ Daniel Etiemble}
\IEEEauthorblockA{\textit{LISN} \\
\textit{University Paris Saclay}\\
Gif sur Yvette, France \\
de@lri.fr}
}

\maketitle

\begin{abstract}
In Carry Propagate Adders, carry propagation is the critical delay. For the 1-digit adders that they use, the most efficient scheme is to generate two intermediate carries: C$_{out0}$ ($C_{in}$=0) and $C_{out1}$($C_{in}$=1). Then multiplex them to produce the correct output according to $C_{in}$. For any radix, the carry output has always a logical value 0 or 1. We show that using 0 and $V_{dd}$ levels for input and output carries instead of 0 and $V_{dd}$/3 in quaternary full adders significantly reduce the carry propagation. We compare such a quaternary full adder with binary full adders to implement N-digit carry propagate adders.
\end{abstract}


\section{Introduction}
\begin{figure}[htbp]
\centerline{\includegraphics[width=8cm]{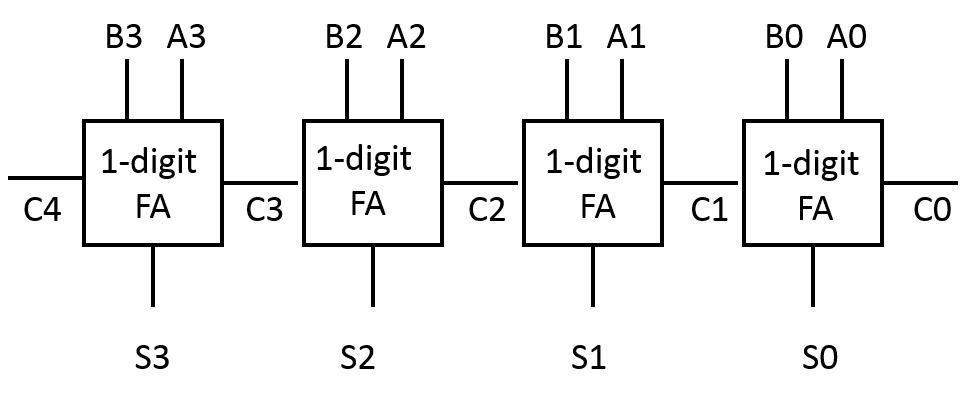}}
\caption{4-digit Carry Propagate Adder}
\label{CPA4}
\end{figure}

Carry Propagate Adders (CPAs) are the most simple N-digit adders. Fig.\ref{CPA4} presents a 4-digit CPA. In this paper, we consider quaternary CPAs. From Fig.\ref{CPA4}, it results that the performance of a CPA is a direct function of the used 1-digit full adder. More precisely, the critical delay path of CPA is related to the carry propagation. 
Several quaternary full adders with CNTFET simulations have been published in the last decade \cite{R1,R2,R3}. 
The truth table of the quaternary full adder is shown in Table \ref{T1}. So, quaternary 1-digit full adders have quaternary inputs and outputs logical values 0,1,2,3 (corresponding to 0,$V_{dd}$/3, 2*$V_{dd}$/3, $V_{dd}$) and carry logical values 0,1 (corresponding generally to $V_{dd}$/3).

\begin{table}
\centering
\caption{Truth table of a quaternary adder}
\begin{tabular}{|c|c|c||c|c|c|c|c|c||c|c|}
  \hline
A&B&Ci&QS&QC& &A&B&Ci&QS&QC\\
\hline
 0&0&0&0&0&&0&0&1&1&0\\
0&1&0&1&0&&0&1&1&2&0\\
0&2&0&2&0&&0&2&1&3&0\\
0&3&0&3&0&&0&3&1&0&1\\

 1&0&0&1&0&& 1&0&1&2&0\\
1&1&0&2&0&&1&1&1&3&0\\
1&2&0&3&0&&1&2&1&0&1\\
1&3&0&0&1&&1&3&1&1&1\\

 2&0&0&2&0&& 2&0&1&3&0\\
2&1&0&3&0&&2&1&1&0&1\\
2&2&0&0&1&&2&2&1&1&1\\
2&3&0&1&1&&2&3&1&2&1\\

 3&0&0&3&0&&3&0&1&0&1\\
3&1&0&0&1&&3&1&1&1&1\\
3&2&0&1&1&&3&2&1&2&1\\
3&3&0&2&1&&3&3&1&3&1\\
  \hline
\end{tabular}
\label {T1}
\end{table}

There are two techniques to get 0, 1, 2 and 3 values. The first one uses three power supplies $V_{dd}$, $2V_{dd}$/3 and $V_{dd}$/3. The second one uses only one power supply ($V_{dd}$) and get the intermediate values through transistors connected as resistors. In that case, there is a large static power dissipation resulting from the direct current flow through the voltage divider. This is why we only consider the option with three power supplies.

The best implementation of ternary adders uses the multiplexer approach \cite {Jaber1}. Similarly, the best implementation to date of quaternary adders (QFAs) \cite{R3} uses this approach. Our implementations are a variant of the QFA proposed in \cite{R3}.
With this approach, all proposed designs use $V_{dd}$/3 as the voltage when carry=1. This  raises one question : do the binary carries must mandatory use 0 and 1 quaternary values or can they use 0 and 3 quaternary values ? In other words, 0 and $V_{dd}/3$ or 0 and $V_{dd}$  when the quaternary adders have a $V_{dd}$ power supply. In this paper, we consider both cases: QFA1 with $V_{dd}/3$ carry swing and QFA2 with $V_{dd}$ carry swing. 

The paper is organized as follow: First, we present the methodology. Then we present two versions QFA1 and QFA2 that uses the MUX-based approach.  Then, we present different 2-bit binary full adders that compute the same amount of information. Finally, we compare the performance of the quaternary and binary full adders and conclude. 

The main contributions of this paper are the following one:
\begin{itemize}
\item We present two versions of a quaternary adders that minimize the carry propagation. With $V_{dd}$ power supply, the first one has $V_{dd}$/3 carry swing and the second one $V_{dd}$ carry swing. The second one has smaller carry delays
\item The two quaternary adders are compared with 2-bit binary ones. Binary adders can use smaller power supplies (such as $V_{dd}$/2) to reduce power dissipation
\item For most features, the 2-bit binary adders are more efficient than the quaternary ones

\end{itemize}

\section{Methodology}
The significant figures to compare circuit designs include switching times, power dissipation, chip area, etc. The comparison is realized by using HSpice simulations and evaluating the chip area according to transistor diameters.
All simulations are done with the 32nm CNTFET parameters of Stanford library \cite{Deng}. We use this technology and these parameters as they are used in most papers presenting ternary or quaternary implementations of adders or multipliers.

Generally, propagation delays are presented as an average of the delays corresponding to all combinations of input transitions. This presentation is confusing. For the CPA presented in Fig.\ref{CPA4}, $A_{i}$, $B_{i}$ (with i=0,1,2,3) and $C_0$ inputs are simultaneously available. The important information is the propagation delay corresponding to the critical paths. For the 4-digit CPA (Fig.\ref{CPA4}), this critical path is from C$_0$ or $A_0$ or $B_0$ $\to$ $C_1$ (worst case) then from $C_1$ $\to$ $C_2$ then from $C_2$ $\to$  $C_3$ and finally from $C_3$ $\to$  $C_4$ or $S_3$ (worst case). We will only present the propagation delays corresponding to the critical paths.

Both power and PDP (Power Delay Product) dissipation directly depend on the duration of the input signals. It is important to use the same input signals for all simulations.

We use a rough evaluation of the chip area by summing the diameters of all the used transistors by each circuit. In this paper, we use the diameter values presented in Table \ref{Diameter}. The choice of appropriate diameter values is mandatory for the threshold detectors and A¹, A², A³ circuits presented in \ref{QAI}. Some other values are needed to provide sufficient fan-out to the quaternary multiplexers.

Many techniques have been proposed to design full adders. We only consider techniques with the following properties:
\begin{itemize}
\item  No static power dissipation.
\item The circuit outputs have full swing.
\item  The circuits should have a sufficient driving capability.
\end{itemize}

\begin{table}
\centering
\caption{Transistor diameter}
\begin{tabular}{|c|c|c|c|c|c|c|c|c|}
  \hline
n&8&10&13&19&29&37\\
\hline
D (nm)&0.626&0.783&1.017&1.487&2.270&2.896\\
  \hline
\end{tabular}
\label {Diameter}
\end{table}

\begin{figure}[htbp]
\centerline{\includegraphics[width=8cm]{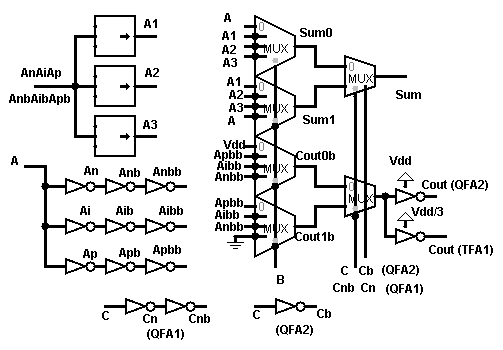}}
\caption{Quaternary Full Adder (MUX approach)}
\label{QFAMUX}
\end{figure}

\section{New Quaternary Full Adders}
The common scheme is presented in Fig. \ref{QFAMUX}. The two QFAs only differ by the carry operation. Carry input and output values are 0, $V_{dd}/3$ (QFA1) and 0, $V_{dd}$ (QFA2). The control of the two MUX2 is shown is Fig. \ref{QFAMUX}. The carry out is obtained by inverters with $V_{dd}/3$ supply (QFA1) or $V_{dd}$ supply (QFA2)

\subsection{Quaternary Adder Implementation} \label{QAI}
\begin{figure}[htbp]
\centerline{\includegraphics  [width =5 cm]{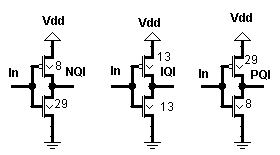}}
\caption{Threshold detectors}
\label{QDEC4}
\end{figure}

\begin{figure}[htbp]
\centerline{\includegraphics[width=8cm]{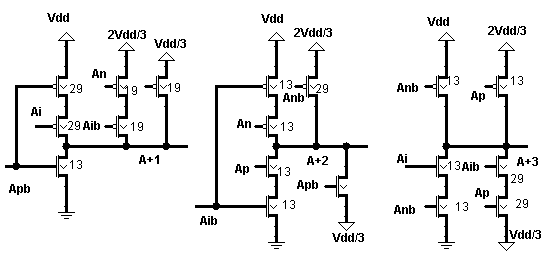}}
\caption{A¹, A² and A³ circuits}
\label{A1A2A3}
\end{figure}

\begin{figure}[htbp]
\centerline{\includegraphics[width=6cm]{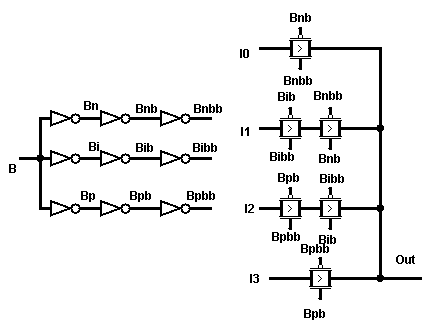}}
\caption{4-input MUX with quaternary control}
\label{QMUXDE}
\end{figure}

The common functional scheme is shown in Fig. \ref{QFAMUX}. The threshold detectors (Fig. \ref{QDEC4}), the circuits A¹, A², A³ (Fig. \ref{A1A2A3}) and the MUX4 (Fig. \ref{QMUXDE}) are similar to those presented in \cite{R3}. The two final multiplexers are typical binary multiplexers. $\overline{Cout}$ is computed from $\overline{C_{out0}}$ and $\overline{C_{out1}}$. A final inverter delivers $C_{out}$. 4-input multiplexers with quaternary control are used (Fig. \ref{QMUXDE}). The three inverters with outputs $B_{nbb}$, $B_{ibb}$ and $B_{pbb}$ are also buffers because inverters $B_{n}$ and $B_{p}$ have poor driving capability. Paper \cite{R3} first uses a quaternary half adder (sum and carry circuits). A second stage computes the final result by adding +1 mod(4) to sum when $C_{in}$=1 and computing $C_{out}$ according to $C_{in}$. We directly computes Sum and $C_{out}$ within a single stage. $C_{in}$ to $C_{out}$ propagation delay is reduced to a MUX2 and final inverter path.

\subsection{Performance}
\begin{figure}[htbp]
\centerline{\includegraphics[width=8cm]{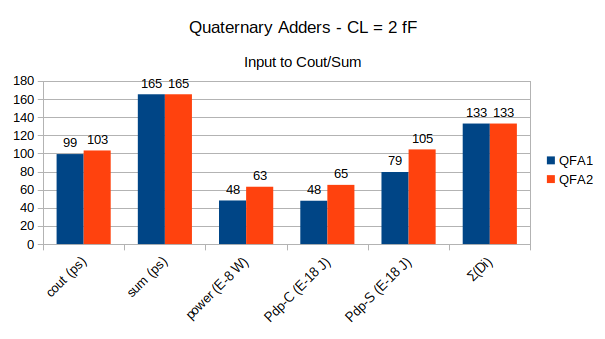}}
\caption{Input to $C_{out}$/Sum performance for QFA1 and QFA2}
\label{QFAI20}
\end{figure}

\begin{figure}[htbp]
\centerline{\includegraphics[width=8cm]{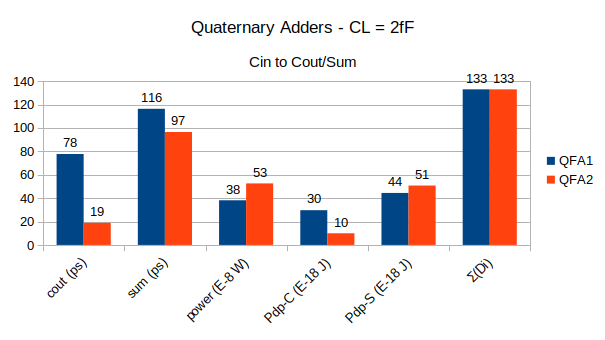}}
\caption{$C_{in}$ to $C_{out}$/Sum performance for QFA1 and QFA2}
\label{QFAC20}
\end{figure}

For all simulations, the same input waveforms are used. Extensive simulations have determined that 0$\to$1$\to$2$\to$3$\to$2$\to$1 $\to$0 for input a with $C_{in}$=0 lead to the input to $C_{out}$/Sum worst case delays. Similarly, 0$\to $1(QFA1) or 3(QFA2) $\to $0 with A=2 and B=1 lead to the $C_{in}$ to $C_{out}$/Sum worst case delays. These configurations are used to evaluate the performance of QFA1 and QFA2. The only difference is the amplitude of the carry swing. 
The performance results are presented in Fig. \ref{QFAI20} and \ref{QFAC20}. These figures provide the data  and allows a direct comparison for each feature. The significant information is Input to $C_{out}$ (first adder of a CPA), $C_{in}$ to $C_{out}$ (following adders) and $C_{in}$ to Sum (last adder of a CPA). Only simulation results with CL = 2fF are presented.

QFA1 and QFA2 have simular $\Sigma{Di}$. QFA1 has a small advantage in term of power. However, it is outperformed by QFA2 for $C_{in}$ to $C_{out}$ delay, which is the critical delay for a CPA. The situation is the same for PDP. This big advantage comes from the last carry inverter that is more efficient with a $V_{dd}$ supply than with a $V_{dd}/3$ supply. 

\section{Binary Full Adders}
\begin{figure}[htbp]
\centerline{\includegraphics[width=8cm]{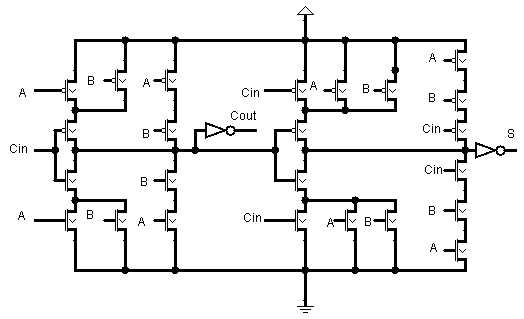}}
\caption{28T Binary Full Adder - BFA1}
\label{28T}
\end{figure}

\begin{figure}[htbp]
\centerline{\includegraphics[width=6cm]{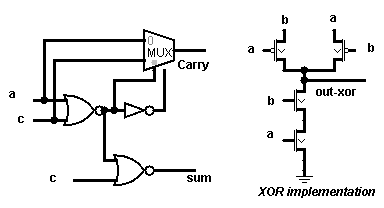}}
\caption{14T Binary Full Adder - BFA2}
\label{14T}
\end{figure}

For comparison, we use two different binary full adders. The first one (BFA1) is the typical 28 T binary full adder (Fig. \ref{28T}). The second one (BFA2) is shown in Fig. \ref{14T}. It uses 14 transistors.
The two binary full adders operate with the same $V_{dd}$=0.9V as the quaternary adder. They can also operate with $V_{dd}$=0.45V, which roughly divide by 4 the dynamic power dissipation. $V_{dd}$=0.45V is a too small power supply value to operate with the four levels of a quaternary adder.

\begin{figure}[htbp]
\centerline{\includegraphics[width=9cm]{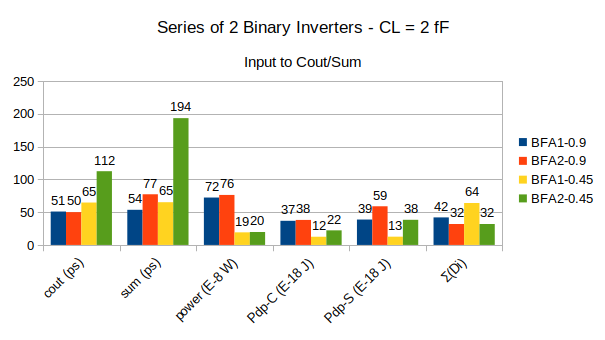}}
\caption{Series of two adders - CL=2 fF - Input to $C_{out}$/Sum}
\label{2BFAI20}
\end{figure}

\begin{figure}[htbp]
\centerline{\includegraphics[width=9cm]{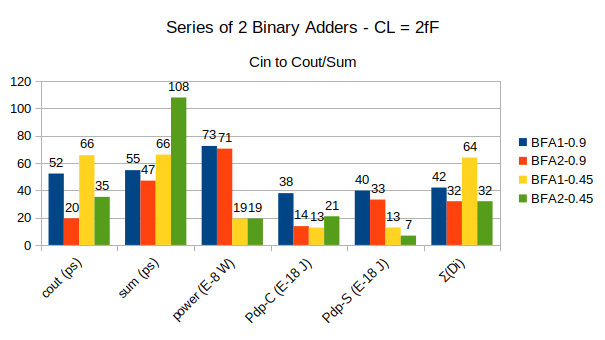}}
\caption{Series of two adders - CL=2 fF -$C_{in}$ to $C_{out}$/Sum}
\label{2BFAC20}
\end{figure}

We simulate a series of two binary adders with $V_{dd}$ = 0.9V and $V_{dd}$ = 0.45V to be compared to the quaternary adder.
Only simulation results with CL = 2fF are presented.
The performance for Input to $C_{out}$/Sum with CL = 2fF are given in Fig \ref{2BFAI20}. 
The performance for $C_{in}$ to $C_{out}$/Sum with CL = 2fF are given in Fig \ref{2BFAC20}. 

\section{Overall comparison}
We compare QFA2 (best QFA) with the two versions of BFA2 (best BFA). The data provided in this paper allow the other comparisons.

\begin{figure}[htbp]
\centerline{\includegraphics[width=9cm]{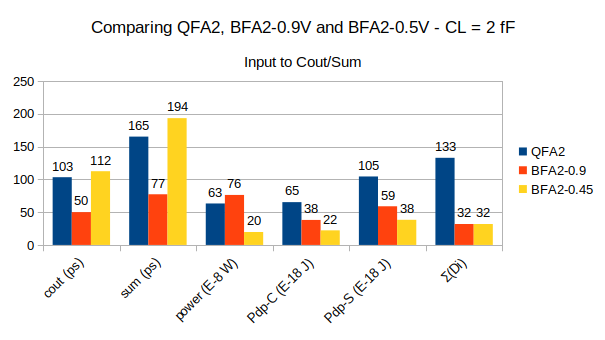}}
\caption{Performance comparison for Input to $C_{out}$/Sum}
\label{CTBI20}
\end{figure}

Fig. \ref{CTBI20} presents the comparison results for Input to $C_{out}$/Sum. For the only significant delay (Input to $C_{out}$), the two versions of the binary adder have better performance. Fig. \ref{2BFAC20} present the comparison results for $C_{in}$ to $C_{out}$/Sum. The quaternary adder has a small advantage for $C_{in}$ to $C_{out}$ (only one stage versus two for the binary ones) and for power versus BFA-0.9V. However, BFA-0.45V is better in terms of power and PDP. The quaternary adder has a huge disadvantage versus BFA2 in terms of $\Sigma{Di}$  with a x4 ratio. 43\% of $\Sigma{Di}$  comes from MUX4s and MUX2s.

As the comparison between quaternary and binary circuits are done with the same technology, there are few opportunities to get different results in the comparisons for different capacitive loads and different temperatures.

\begin{figure}[htbp]
\centerline{\includegraphics[width=9cm]{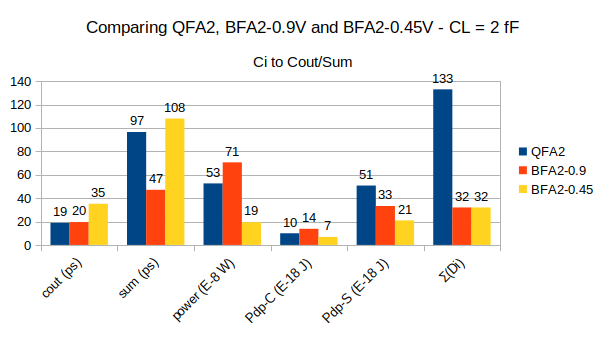}}
\caption{Performance comparison for $C_{in}$ to $C_{out}$/Sum}
\label{CTBC20}
\end{figure}

\section{Concluding remarks}
The best implementation of quaternary full adders should minimize the $C_{in}$ to $C_{out}$ propagation delay. We have presented two versions of a quaternary full adder for which the $C_{in}$ to $C_{out}$ delay is reduced to the path through one MUX2 and one inverter. The first one has the typical 0 and $V_{dd}$/3 carry levels while the second one uses a $V_{dd}$ carry swing. For CPAs, the QFA2 version is more efficient. 

We have compared the new quaternary adders with two binary adders. For carry propagation and power, QFA2 is better than BFA2 using the same 0.9V power supply. BFA2 with 0.45V power supply has a significant advantage in terms of power and PDP. The quaternary full adder uses far more chip area. It mainly comes from the needed MUX4 circuits. Inverters used as buffers are also needed to overcome the driving capability weakness of the QN and QP circuits.

CPAs are circuits for which moving from binary to quaternary N-digit CPAs is simple: just replace the binary full adders by quaternary full adders. They are among the most favorable circuits for the quaternary approach. Moving from a 2N*2N-bit binary multiplier to a N*N digit quaternary multiplier is not so simple as quaternary multipliers uses both quaternary 1-digit adders and 1-digit multipliers.

\end{document}